\documentclass[sigconf, anonymous=false]{acmart}

\AtBeginDocument{%
  }

\copyrightyear{2023}
\acmYear{2023}
\setcopyright{none}
\acmConference[ARES 2023]{The 18th International Conference on Availability, Reliability and Security}{August 29-September 1, 2023}{Benevento, Italy}
\acmBooktitle{The 18th International Conference on Availability, Reliability and Security (ARES 2023), August 29-September 1, 2023, Benevento, Italy}
\acmPrice{15.00}
\acmDOI{10.1145/3600160.3605011}
\acmISBN{979-8-4007-0772-8/23/08}

\usepackage[nolist]{acronym}
\usepackage{amsfonts}
\usepackage[capitalize]{cleveref}
\usepackage{graphicx}
\usepackage{siunitx}
\usepackage{subcaption}
\usepackage{url}

\begin{acronym}
  \acro{API}{Application Programming Interface}
  \acro{BLE}{Bluetooth Low Energy}
  \acro{CTA}{Contact Tracing App}
  \acro{COVID-19}{coronavirus disease 19}
  \acro{CWA}{Corona-Warn-App}
  \acro{GAEN}{Google/Apple Exposure Notification}
  \acro{GDPR}{General Data Protection Regulation}
  \acro{IBL}{Inter-Broadcast Latency}
  \acro{PIA}{Privacy Impact Assessment}
  \acro{TLS}{Transport Layer Security}
  \acro{UUID}{Universally Unique Identifier}
  \acro{OS}{Operating System}
  \acro{W3C}{World Wide Web Consortium}
\end{acronym}

\makeatletter%
\providecommand\color[2][]{%
\renewcommand\color[2][]{}%
}%
\providecommand\transparent[1]{%
\renewcommand\transparent[1]{}%
}%
\providecommand\rotatebox[2]{#2}%
\newcommand*\fsize{\dimexpr\f@size pt\relax}%
\newcommand*\lineheight[1]{\fontsize{\fsize}{#1\fsize}\selectfont}%
\ifx\svgwidth\undefined%
\ifx\svgscale\undefined%
  \relax%
\else%
\fi%
\else%
\fi%
\global\let\svgwidth\undefined%
\global\let\svgscale\undefined%
\makeatother%

\begin{document}

\title[Feasibility Study on Fingerprinting the Corona-Warn-App]{Smartphones in a Microwave: Formal and Experimental Feasibility Study on Fingerprinting the Corona-Warn-App}

\author{Henrik Graßhoff}
\affiliation{%
\institution{University of Münster}
\city{Münster}
\country{Germany}
}
\email{grasshoff@uni-muenster.de}

\author{Florian Adamsky}
\affiliation{%
\institution{Hof University of Applied Sciences, Institute of Information Systems}
\city{Hof}
\country{Germany}
}
\email{florian.adamsky@hof-university.de}

\author{Stefan Schiffner}
\affiliation{%
\institution{BHH University of Applied Sciences}
\city{Hamburg}
\country{Germany}
}
\email{stefan.schiffner@bhh.hamburg.de}

\renewcommand{\shortauthors}{Graßhoff et al.}

\begin{abstract}
  \acp{CTA} have been developed to contain the \ac{COVID-19} spread.
  By design, such apps invade their users' privacy by recording data about their health, contacts, and---partially---location.
  Many \acp{CTA} frequently broadcast pseudorandom numbers via Bluetooth to detect encounters.
  These numbers are changed regularly to prevent individual smartphones from being trivially trackable.
  However, the effectiveness of this procedure has been little studied.
  
  We measured real smartphones and observed that the German \ac{CWA} exhibits a device-specific latency between two subsequent broadcasts.
  These timing differences provide a potential attack vector for fingerprinting smartphones by passively recording Bluetooth messages.
  This could conceivably lead to the tracking of users' trajectories and, ultimately, the re-identification of users.
\end{abstract}




\maketitle

\acresetall

\section{Introduction}

The coronavirus pandemic was the first pandemic in which we, i.e. humanity, had the means to observe its spread in real time. This wealth of information posed and continues to pose a challenge to societies around the world.
More information allows to take the right decisions to slow the spread of a pandemic, one might conclude.
Or does it though?

The greater goal, limiting the spread of the virus or at least slowing it down, is in conflict with individual freedom rights.
In a first response, many governments opted to bring public live to a halt.
This understandable and early response was not sustainable.
Traditionally, one would aim to only isolate those who are infected, but this approach was undermined by the fact that individuals with asymptomatic infection can also transmit the virus~\cite{he2020temporal}.

This has led to the first ever large-scale introduction of automatic contact tracing by means of \acp{CTA}.
Such apps record their users' contacts and alert them in case of a close encounter with an infected person.
In 2020, Google and Apple integrated extensive contact tracing functionality into their respective mobile operating systems, and many national authorities worldwide deployed \acp{CTA} since then that have been used by millions of people, cf.~\cite{Prodan_Birov_2022} for more information on downloads and active usage in Europe.

\acp{CTA} inherently concern their users' privacy as they process personal contact and health data.
The German \ac{CWA} \cite{open-source-project-cwa} and numerous other \acp{CTA} operate by broadcasting a pseudorandom number (\emph{pseudonym}) several times per second via \ac{BLE} to all nearby devices.
Linking the pseudonym to a real person might allow an adversary to gain insights into their infection status or movement patterns.
Developers have implemented basic privacy protection mechanism, but their effectiveness has not been proven.
Due to this conceivable profound privacy threat, many legal frameworks, particularly the \ac{GDPR}~\cite{gdpr}, hence require a \ac{PIA}, which must be based on a thorough threat analysis.
This \ac{PIA} should happen in the light of article~9 \ac{GDPR} which establishes a special protection of health data.

As the virus' spread slowed down and with the wide availability of vaccines, many \acp{CTA} are discontinued, i.e. infrastructure has been scaled down or switched off and maintenance of the apps has been brought to a hold.
Therefore, imitate privacy risks of \acp{CTA} got reduced.
However, two concerns remain:
First, what happens to the contact tracing functionality implemented in the Android and iOS operating systems?
Has this come to stay and pose a continuous threat to security?
Second, while maintenance for e.g. the \ac{CWA} has stopped, it is not actively removed from users' phones but considered to hibernate.\footnote{The German Federal Minister of Health, Karl Lauterbach, says that as of June the German \ac{CWA} will hibernate (\href{https://www.tagesschau.de/inland/innenpolitik/corona-warn-app-ende-100.html}{https://www.tagesschau.de/inland/innenpolitik/corona-warn-app-ende-100.html}).}
However, the semantics are unclear in many ways:
Under which circumstances can such a hibernating app be woken up, i.e. which political and scientific process decides if a new pandemic is severe enough?
What form of maintenance will be provided for such an app?

While the above questions are not subject of this paper, we observe that WHO epidemiologists expect that ``COVID-19 will not be the last'' pathogen with pandemic potential and the next one ``could appear at any time'' \cite{van2021preparing}.
With this paper, we aim to contribute to the \ac{PIA} if electronically aided contact tracing is re-considered in the future.

Our contribution consists of two experiments.
We used low-cost and off-the-shelf hardware to monitor the \ac{BLE} sending behavior of smartphones with the German \ac{CWA} installed.
In our first experiment, we observed 15~smartphones in a shielded laboratory environment.
It showed that the average latency between two successive broadcasts varies across devices and is stable over time.
This characteristic acted as a fingerprint for some device and uniquely identified them among all tested phones.
We were able to replicate our observations in a second experiment in busy public places in the city of Münster.
To the best of our knowledge, our paper provides the first study investigating device fingerprinting of smartphones running a \ac{CTA}.
This in turns brings us to the conclusion that further investigations are needed.

\section{Related Work}

A large body of research exists on fingerprinting computing devices.
Publications on fingerprinting typically fall into two categories: logical fingerprints and physical fingerprints.
In the first case, devices are distinguishable due to differences in their software behavior; in the latter case, devices are different due to some physical process, e.g. manufacturing tolerances of a crystal which in turn influences the exact clock rate of a device.

\paragraph{Fingerprinting on Logical Behavior.}
Browser Fingerprinting aims to create fingerprints of web browsers to recognize returning visitors to a website.
In 2009, Mayer~\cite{anonymity:mayer:thesis:2009} conducted a small-scale experiment and collected different information such as \textsc{JavaScript} objects (e.g. \texttt{navigator}, \texttt{screen}, \texttt{Plugin}, \texttt{MimeType}, among others) from \num{1328} web browser to generate a fingerprint.

\emph{Panopticlick}~\cite{panopticlick:eckersley:inproceedings:2010} replicated and extended the former results in 2010 in a large-scale experiment with \num{470162} browser fingerprints and additional features with Flash and Java.
These studies marked the beginning of a discipline; since then, the scientific community has improved fingerprinting continuously, further aided by the introduction of new \acp{API} by the \ac{W3C} to provide rich multimedia content on web pages.
Studies~\cite{mowery_pixel_2012,acar_web_2014} discovered that the Canvas \ac{API} could be exploited to offer high-entropy attributes for a fingerprint.
Further, a study~\cite{webgl:cao:2017:inproceedings} designed fingerprinting techniques based on the WebGL \ac{API}.
We refer interested readers to \cite{browser-fp-survey:laperdrix:article:2019} for a detailed survey of browser fingerprinting.

Researchers~\cite{280012,frolov_use_2019,husak_https_2016} found that even complex network protocols such as \ac{TLS} and OpenVPN are fingerprintable by the protocol handshake.
Similarly for Bluetooth, Celosia and Cunche~\cite{celosia2019fingerprinting} showed that the GATT profile of the Bluetooth stack contains identifying characteristics.
By connecting to nearby discoverable devices, they could collect complete GATT profiles to obtain fingerprints which are unique in many cases.

\paragraph{Fingerprinting Using Physical Attributes.}
Crystal oscillators are being used to generate the required frequency for any radio device.
Due to small imperfections in production, their actual frequencies are slightly off target~\cite{leen_mimo_2017}; hence, devices have a unique frequency.
It has been shown that this frequency offset can be used to distinguish devices~\cite{csi:adamsky:inproceeding:2018,csi:hua:inproceeding:2018}.
Similar results have been established using the deviation of the device clock's speed from real-time \cite{jana2008fast,kohno2005remote}.
For Bluetooth, Huang et al.~\cite{huang2014blueid} exploited the frequency hopping behavior to extract a device's \emph{clock skew} and use this as a fingerprint.

~\\
Our work falls between these two broad categories:
We measure timing behavior which is partially influenced by the logic of the \ac{CWA}, the logic of the underlying \ac{API} by Google and Apple, and the logic of the operating system and particularly the \ac{BLE} stack, but at the same time, our measurements are influenced by the accuracy of the underlying clocks.

\section{Technical Background}

This section explains the technical foundation of the \ac{CWA} which uses \acl{BLE} for broadcasting the pseudonyms provided by the \ac{GAEN} \ac{API}.

\subsection{Bluetooth Low Energy}

\ac{BLE}~\cite{bluetooth} is a wireless communication standard introduced in 2010.
Initially designed for battery-powered gadgets such as smartwatches and Internet of Things applications, it is nowadays supported by almost all modern devices.
\ac{BLE} uses 40~channels in the \SI{2.4}{\giga\hertz} ISM~band;
37 of these are used for data transfer while the other three \emph{advertising channels} are reserved for devices to signal their presence.
To do so, a device broadcasts \emph{advertisement} frames to all nearby devices indicating e.g. its connectibility and characteristics.
Moreover, this broadcast mechanism can be used to transmit small amout of data without establishing a connection between sending and receiving device.

The sender of a \ac{BLE} message is identified by a 48~bit MAC~address.
For basic privacy protection, \ac{BLE} introduced randomized MAC~addresses: instead of broadcasting its globally unique MAC~address, the device can generate a random number to be sent in place of the persistent identifier.
Due to the length of this number, collissions occur extremely rarely so that the randomized MAC~address is a unique identifier for its period of validity.
The longevity and change is carried out by the device, the Bluetooth specification \cite[Vol.~3, Part~C, App.~A]{bluetooth} merely recommends to change it after at most 15~minutes.

\begin{figure*}
	\renewcommand{\arraystretch}{0.4}
	\centering
	\sffamily
	\small

  \begin{subfigure}[t]{.35\textwidth}
    \centering
    \begin{tabular}{|c|c|c|}
      \hline
      \multicolumn{3}{|c|}{\rule{0pt}{.2cm}} \\
      \multicolumn{3}{|c|}{\textbf{Protocol Data Unit of a \ac{BLE} advertisement}} \\
      \multicolumn{3}{|c|}{\rule{0pt}{.2cm}} \\
      \hline
      \multicolumn{3}{c}{} \\
      \hline

      & \multicolumn{2}{c|}{} \\
      Header & \multicolumn{2}{c|}{Payload} \\
      (\SI{2}{\byte}) & \multicolumn{2}{c|}{(\SIrange{6}{37}{\byte})} \\
      & \multicolumn{2}{c|}{} \\

      \hline
      \multicolumn{3}{c}{} \\
      \cline{2-3}

      \multicolumn{1}{c|}{} && \\
      \multicolumn{1}{c|}{} & AdvA & AdvData \\
      \multicolumn{1}{c|}{} & (\SI{6}{\byte}) & (\SIrange{0}{31}{\byte}) \\
      \multicolumn{1}{c|}{} && \\
      \cline{2-3}
    \end{tabular}
    \caption{}
    \label{fig: BLE}
  \end{subfigure}
  ~
  \begin{subfigure}[t]{.65\textwidth}
    \centering
		\begin{tabular}{|c|c|c|c|}
            \hline
    		\multicolumn{4}{|c|}{\rule{0pt}{.2cm}} \\
    		\multicolumn{4}{|c|}{\textbf{AdvData (\ac{GAEN})}} \\
    		\multicolumn{4}{|c|}{\rule{0pt}{.2cm}} \\
    
    		\hline
    		\multicolumn{4}{c}{} \\
    		\hline
    
    		&& \multicolumn{2}{c|}{} \\
    		Flags & Service \ac{UUID} 0xFD6F & \multicolumn{2}{c|}{Data} \\
    		(\SI{3}{\byte}) & (\SI{4}{\byte}) & \multicolumn{2}{c|}{(24\,B)} \\
    		&& \multicolumn{2}{c|}{} \\
    
            \hline
    		\multicolumn{4}{c}{} \\
    		\cline{3-3}
    
    		\multicolumn{2}{c|}{} && \multicolumn{1}{c}{} \\
    		\multicolumn{2}{c|}{} & Pseudonym & \multicolumn{1}{c}{$\dotsb$} \\
    		\multicolumn{2}{c|}{} & (\SI{16}{\byte}) & \multicolumn{1}{c}{(\SI{8}{\byte})} \\
    		\multicolumn{2}{c|}{} && \multicolumn{1}{c}{} \\
    		\cline{3-3}
    	\end{tabular}
    \caption{}
    \label{fig: GAEN}
    \end{subfigure}

	\caption{Structure and content of (\subref{fig: BLE}) the Protocol Data Unit of a \texttt{ADV\_NONCONN\_IND} type \ac{BLE} advertisement \cite[Vol.~6, Part~B, Sec.~2.3]{bluetooth} and (\subref{fig: GAEN}) the \texttt{AdvData} field of a \ac{GAEN} broadcast \cite{gaen-bluetooth}}
	\label{fig: broadcast}
\end{figure*}

\subsection{Exposure Notification}

In April 2020, Apple and Google jointly announced the integration of contact tracing directly into their respective mobile \ac{OS}, naming it \emph{Exposure Notification}~\cite{gaen}, or \ac{GAEN} for short.
When activated by the user, the \ac{OS} generates pseudorandom 128~bit numbers (\emph{pseudonyms}) which are changed every \SIrange{10}{20}{\minute} according to the documentation \cite{gaen-bluetooth}.
\ac{GAEN} frequently emits these pseudonyms with a recommended waiting time of \SIrange{200}{270}{\milli\second} between two sendings, a delay which we refer to as \ac{IBL}.
Additionally, the device listens to other smartphones' broadcasts and logs the pseudonyms it receives.
An infected individual can decide to upload specific keys to a server which allow other phones to reconstruct their emitted pseudonyms.
This key material is regularly fetched by every participating smartphones and employed to calculate a contagion risk for its user.

This functionality is implemented as an \ac{API} on the \ac{OS} level.
Authorized apps like the \ac{CWA} can access this \ac{API} to provide a frontend to the user, but the underlying contact tracing function---especially the \ac{BLE} broadcasting---is nevertheless not influenced by the app.

The \ac{GAEN} \ac{API} emits its information in the data unit of an \texttt{ADV\_NONCONN\_IND} type advertisement packet whose general structure is shown in \cref{fig: BLE}.
It is comprised of a 2 byte header and a payload of variable size.
The latter contains a field \texttt{AdvA} for the device's (possibly randomized) MAC~address as well as the \texttt{AdvData} whose content is variable.
As can be seen in \cref{fig: GAEN}, Google and Apple have defined it to contain two main ingredients:
\begin{itemize}
  \item A \ac{UUID} 0xFD6F by which other smartphones can detect a \ac{GAEN} broadcast among packets for other purposes.
  \item The pseudorandom 16~bit contact tracing pseudonym.
\end{itemize}
The MAC~address is randomized in sync with the pseudonym as performing their change asynchronously would clearly annihilate the intended privacy protection.

The present, very frequent broadcasting is a design decision in favor of the app's utility.
While a lower broadcasting frequency would restrict the possibility of continuous device monitoring, it would also increase the risk of not detecting infectious encounters, eventually making the app less valuable from a medical perspective.

\section{Formal Background and Privacy Metric}

In this section, we provide the terminological and mathematical background of analysing fingerprintability.

\subsection{Pseudonym Types and Anonymity}

In general, pseudonyms are identifying an entity in a given context. If the relation between the identity and pseudonym can be hidden from an adversary, a pseudonym can provide a certain level of privacy protection.
The level of privacy protection pseudonyms can provide is depending on their usage; in particular, it depends on for how long and in which contexts they are used.
For a systematic discussion on pseudonym types we refer to \cite{pfitzmann2010terminology}.

Following this terminology, \ac{GAEN} pseudonyms are short term role pseudonyms, i.e. in its role as a participant of \ac{GAEN}, an app user provides this pseudonym during interactions with other app users for a given time period.
These pseudonyms are only used in the context of the app, hence by individuals in their role as app users.
In the case of \ac{GAEN}, users are broadcasting their pseudonym for \SIrange{10}{20}{\minute}.
During said period, all broadcast messages of the same user can be linked to each other.
After said time period, users will change their pseudonym, rendering it theoretically impossible to trivially \emph{link} pseudonyms of different time periods.
Here by \emph{link} we mean that an adversary can distinguish if two or more pseudonyms belong to the same entity or not.

The way pseudonyms are used in \ac{GAEN} allows a certain level of privacy protection: under the assumption that pseudonyms of different time periods cannot be \emph{linked}, users' trajectories cannot be reconstructed even if an adversary can observe broadcast messages at many locations and over longer periods of time. In other words, these pseudonyms provide a certain level of conditional anonymity.

\subsection{Mathematical Treatment}
\label{subsec: privacy metric}

To measure fingerprintability, we adapt the \emph{degree of anonymity} model proposed by Diaz et al.~\cite{diaz2002towards}.
The authors consider an adversary whose goal is to deanonymize the users of a system, e.g. a sender-recipient system.
By observing the system, the adversary obtains probabilities about whether a user is the sender of a particular message.
The normalized Shannon entropy of this probability distribution is then taken as a measure for the anonymity that the system provides.

Transferring this to fingerprinting, suppose an adversary observes $n$ data points $X = \{x_1,\ldots,x_n\} \subseteq \mathbb{R}$ over time originating from different entities and tries to group the data points according to their sources.
For each entity, the data may vary and therefore can only be measured with some uncertainty $\varepsilon > 0$ even if the adversary has unrestricted measuring accuracy.
A fingerprinting attack then is the attempt to partition $X$ into subsets of data originated from the same entity.
Such an attack is obviously more successful if the data are precise (i.e. $\varepsilon$ is small) and admit high variation.

The amount of information that the adversary gains from the observed characteristic $X$ can be quantified as follows:
By grouping the data set $X$ into $k$ bins $1,\ldots,k$ of width $\varepsilon$, we obtain the histogram of a discrete probability distribution.
The probability $p_i$ of bin $i=1,\dotsc,k$ is given by the number of elements in that particular bin divided by the number of total data points $n$.
Practically, elements in the same bin can be considered indistinguishable by the adversary as their distance is at most the uncertainty $\varepsilon$.
Hence, the maximum information the adversary can obtain from their observations is quantified by the Shannon entropy of that histogram:
\begin{equation}
  \tag{where $0\log_2(0)=0$}
  H(X) = - \sum_{i=1}^k p_i \log_2(p_i)
\end{equation}

Technically, note that the probabilities $p_i$ depend not only on $\varepsilon$ but on the location of the bins on the $x$-axis as well which we did not define.
The above term $H(X)$ is understood to be the maximum of the right hand side over all (finitely many) probability distributions for different bin locations.
Even an adversary with unlimited background knowledge could not gain more than $H(X)$ information from their observations.

If data precision and variation are high, then $k > n$ and $p$ is the uniform distribution $p_i = \frac{1}{n}$ which results in a maximum entropy of $\log_2(n)$.
On the other hand, a low precision or variation leads to data points from different entities in the same bin and in the most extreme case of $p_i = 1$ for one bin $i$ to $H(X) = 0$.

Similar to~\cite{diaz2002towards} we say that the data set $X$ with precision $\varepsilon$ provides a \emph{fingerprinting anonymity} of
\[ A(X,\varepsilon) = 1-\frac{H(X)}{\log_2(n)} \in [0,1]. \]
Note that our definition is almost literally the same as the \emph{degree of anonymity} given in~\cite{diaz2002towards}, but in our case, the attacker knowledge is represented by the histogram entropy $H(X)$ instead of $\log_2(n)-H(X)$.
The fingerprinting anonymity reaches its minimum and maximum if
\[
  \arraycolsep=1.6pt
  \begin{array}{lllll}
    A(X, \varepsilon) = 0 & \Leftrightarrow & H(X) = \log_2(n) & \Leftrightarrow & \text{high precision and variation,} \\
    A(X, \varepsilon) = 1 & \Leftrightarrow & H(X) = 0 & \Leftrightarrow & \text{low precision or variation.}
  \end{array}
\]

\section{Experimental Methodology}

This section presents the results of our two performed experiments.
The first experiment was conducted on a small scale in an isolated environment: we collected temporal broadcasting data of 15~smartphones which had the \ac{CWA} installed and running.
After finding device-specific differences in the \acp{IBL}, we proceeded in a second experiment and measured smartphones in multiple public places.
This yielded insights into the \ac{IBL} distribution across an estimated 121~smartphones.
We used this distribution data to evaluate the privacy breach of the \ac{IBL} differences in terms of the fingerprinting anonymity.

\subsection{Software and Hardware Setup}

Processing \ac{BLE} broadcasts is feasible with little programming expertise and cheap hardware.
All our measurements were performed using a Python script which operates as follows: collect \ac{BLE} advertisements every \SI{50}{\milli\second} and filter \ac{GAEN} broadcasts by their \ac{UUID} 0xFD6F; group incoming \ac{BLE} broadcasts by their MAC~address as such advertisements originate from the same device; for each device, calculate the latencies between its successive broadcasts and store those between \SI{220}{\milli\second} and~\SI{350}{\milli\second}.

The decision to group \ac{BLE} broadcasts by their MAC~address instead of their \ac{GAEN} pseudonym was made in order to process as little personal data as possible: in contrast to the MAC~addresses, the pseudonyms could potentially leak the Covid infection status of a participant at a later time.
Since MAC~address and pseudonym are changed in sync, both identify a broadcast's source equally well.

As for Bluetooth receiving hardware, we used a Lenovo Ideapad~510S laptop running Fedora Linux.
However, we subsequently verified that the measurements could be carried out identically on a Raspberry Pi~4B with \SI{4}{\giga\byte} of RAM (cf.~\cref{fig: screenshot}).
The attack is thus feasible without significant hardware requirements.

The above methodology was applied in two experiments:

\begin{figure}[t]
  \centering
	\includegraphics[width=\linewidth]{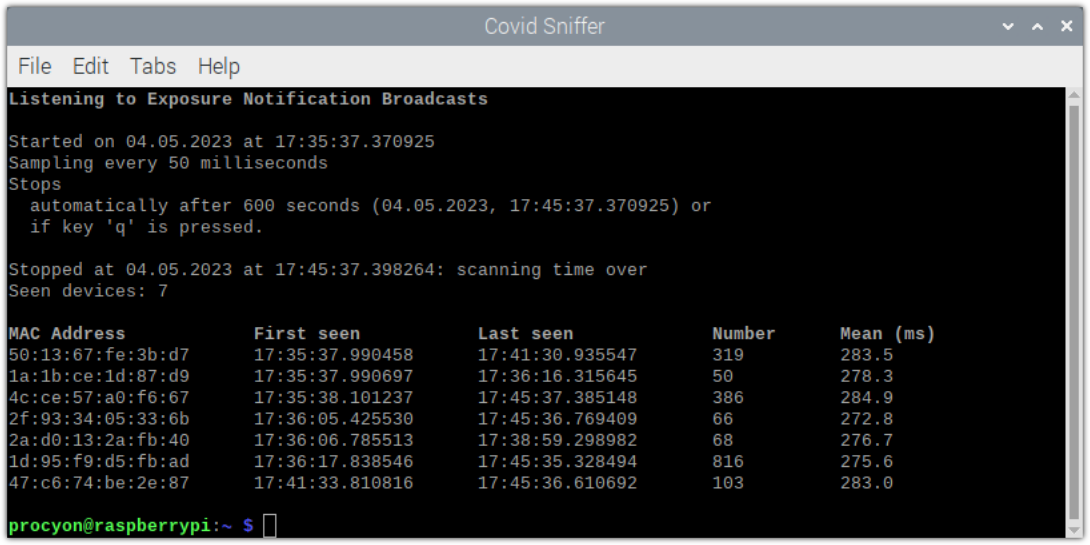}
  \caption{A screenshot of the script running for ten minutes on a Raspberry Pi~4B}
	\label{fig: screenshot}
\end{figure}

\subsection{Laboratory Experiment}

In the first experiment, we measured the \ac{IBL} of the 15~smartphones in \cref{table} in an isolated environment.
At the time of testing, all phones were personal devices in everyday use, meaning that a variety of apps other than the \ac{CWA} were installed, custom settings were made, and some phones could be measured for a longer time and contribute a greater number of pseudonym cycles than others.

While being measured, the phones did not perform any resource-intensive tasks.
Moreover, we isolated the phone and receiver in a common microwave to reduce the influence of other environmental Bluetooth devices.
Considering the full ISM~band, a microwave oven is not a Faraday cage.
It still blocks \SI{2.4}{\giga\hertz}~RF communication sufficiently which is the relevant frequency range for our experiment.
We were able to verify the effectiveness of our isolation by observing that the Python script recorded only a single \ac{BLE} source once the microwave door was closed.

\subsection{Field Experiment}
In the second experiment, we collected the \ac{IBL} of unknown smartphones carried around by present people in public places.
We conducted this experiment in multiple spots in Münster, Germany, in April, July, and August~2022.
Since a pseudonym change could result in a smartphone contributing twice to our data, we limited each measurement to ten minutes.
Subsequently, we rejected entries with less than $50$ data points.

\section{Results}

This section is divided into two parts.
We begin by presenting the main qualitative observations we made in the two experiments.
Afterwards, we evaluate the fingerprinting information leakage by the \ac{IBL} in terms of the privacy metric introduced in \cref{subsec: privacy metric}.

\subsection{Key Findings}

The \ac{IBL} data collected in the laboratory experiment are presented in \cref{table}.
For each device, the \acp{IBL} of a pseudonym cycle were averaged to give the \ac{IBL} mean for this particular pseudonym.
The columns \emph{mean} and \emph{double standard deviation} were then derived from these values.
Hence, when talking about a phone's overall \emph{\ac{IBL} (mean)}, we refer to the average of its pseudonyms means.

\begin{figure}
    \centering
    \begingroup%
    \begin{picture}(1,0.31730327)%
    \lineheight{1}%
    \setlength\tabcolsep{0pt}%
    \put(0,0){\includegraphics[width=\unitlength,page=1]{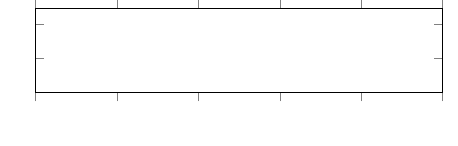}}%
    \put(0.04927053,0.05593474){\color[rgb]{0,0,0}\makebox(0,0)[lt]{\lineheight{1.25}\smash{\begin{tabular}[t]{l}250\end{tabular}}}}%
    \put(0.22717153,0.05593474){\color[rgb]{0,0,0}\makebox(0,0)[lt]{\lineheight{1.25}\smash{\begin{tabular}[t]{l}260\end{tabular}}}}%
    \put(0.40507708,0.05593474){\color[rgb]{0,0,0}\makebox(0,0)[lt]{\lineheight{1.25}\smash{\begin{tabular}[t]{l}270\end{tabular}}}}%
    \put(0.58297808,0.05593474){\color[rgb]{0,0,0}\makebox(0,0)[lt]{\lineheight{1.25}\smash{\begin{tabular}[t]{l}280\end{tabular}}}}%
    \put(0.76087907,0.05593474){\color[rgb]{0,0,0}\makebox(0,0)[lt]{\lineheight{1.25}\smash{\begin{tabular}[t]{l}290\end{tabular}}}}%
    \put(0.93878462,0.05593474){\color[rgb]{0,0,0}\makebox(0,0)[lt]{\lineheight{1.25}\smash{\begin{tabular}[t]{l}300\end{tabular}}}}%
    \put(0.04424968,0.10256536){\color[rgb]{0,0,0}\makebox(0,0)[lt]{\lineheight{1.25}\smash{\begin{tabular}[t]{l}0\end{tabular}}}}%
    \put(-0.00269047,0.17702687){\color[rgb]{0,0,0}\makebox(0,0)[lt]{\lineheight{1.25}\smash{\begin{tabular}[t]{l}0.05\end{tabular}}}}%
    \put(0.01629134,0.25148382){\color[rgb]{0,0,0}\makebox(0,0)[lt]{\lineheight{1.25}\smash{\begin{tabular}[t]{l}0.1\end{tabular}}}}%
    \put(0,0){\includegraphics[width=\unitlength,page=2]{data/OnePlus-Nords.pdf}}%
    \put(0.44278579,0.00057796){\color[rgb]{0,0,0}\makebox(0,0)[lt]{\lineheight{1.25}\smash{\begin{tabular}[t]{l}IBL in ms\end{tabular}}}}%
    \end{picture}%
    \endgroup
  \caption{Distribution of all \ac{IBL} data of the OnePlus Nord~2 (blue, left, 3598 data points) and the OnePlus Nord (orange, right, 17681 data points)}
  \vspace{.6cm}
  \label{fig: two devices}
\end{figure}

\begin{enumerate}
  \item The \ac{IBL} distribution can vary strongly between different devices. \label{obs1}
\end{enumerate}
For example, \cref{fig: two devices} plots the \ac{IBL} distributions of the OnePlus Nord and the OnePlus Nord~2.
Both distributions have evidently little intersection and are separable by a visual inspection with the naked eye.
More comprehensively, the means in \cref{table} range from roughly \SIrange{262}{286}{\milli\second} among all observed devices.
While some smartphones in our test set (such as the Huawei P10 Lite) are uniquely indentifiable by this characteristic, others share a similar \ac{IBL} (e.g. all iPhone~13 Mini or Huawei Mate~10 \& Samsung Galaxy~J7).
We will discuss possible influences on this attribute later in \cref{sec: discussion}.

\begin{figure}
  \centering
  \begingroup%
  \begin{picture}(1,0.41200651)%
    \lineheight{1}%
    \setlength\tabcolsep{0pt}%
    \put(0,0){\includegraphics[width=\unitlength,page=1]{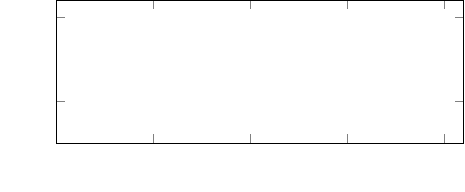}}%
    \put(0.32127256,0.06422231){\color[rgb]{0,0,0}\makebox(0,0)[lt]{\lineheight{1.25}\smash{\begin{tabular}[t]{l}5\end{tabular}}}}%
    \put(0.52080773,0.06422231){\color[rgb]{0,0,0}\makebox(0,0)[lt]{\lineheight{1.25}\smash{\begin{tabular}[t]{l}10\end{tabular}}}}%
    \put(0.72969962,0.06422231){\color[rgb]{0,0,0}\makebox(0,0)[lt]{\lineheight{1.25}\smash{\begin{tabular}[t]{l}15\end{tabular}}}}%
    \put(0.938596,0.06422231){\color[rgb]{0,0,0}\makebox(0,0)[lt]{\lineheight{1.25}\smash{\begin{tabular}[t]{l}20\end{tabular}}}}%
    \put(-0.00294852,0.18167966){\color[rgb]{0,0,0}\makebox(0,0)[lt]{\lineheight{1.25}\smash{\begin{tabular}[t]{l}280\,ms\end{tabular}}}}%
    \put(-0.00294852,0.36273486){\color[rgb]{0,0,0}\makebox(0,0)[lt]{\lineheight{1.25}\smash{\begin{tabular}[t]{l}290\,ms\end{tabular}}}}%
    \put(0,0){\includegraphics[width=\unitlength,page=2]{data/Huawei-Mate-10.pdf}}%
    \put(0.2515641,0.00837383){\color[rgb]{0,0,0}\makebox(0,0)[lt]{\lineheight{1.25}\smash{\begin{tabular}[t]{l}Chronologically ordered pseudonyms\end{tabular}}}}%
  \end{picture}%
  \endgroup
  \caption{Boxplots of the \ac{IBL} for the first 20 observed pseudonyms of the Huawei Mate 10. The blue horizontal line indicates its overall \ac{IBL} mean of 283.04\,ms. The data contain outliers which are not present in the plot due to an appropriate choice of the y-axis range.}
  \vspace{.6cm}
  \label{fig: boxplots}
\end{figure}

\begin{enumerate}
  \setcounter{enumi}{1}
  \item For each device, the \ac{IBL} mean varies little between different pseudonym cycles. \label{obs2}
\end{enumerate}
The rather small standard deviations in \cref{table} indicate little variation of the \ac{IBL} mean between pseudonym cycles.
For instance, the \ac{IBL} means per pseudonym in \cref{fig: boxplots} narrowly fluctuate around the Huawei Mate 10's overall \ac{IBL} mean of \SI{283.04}{\milli\second}.

\begin{figure}
  \centering
  \begingroup%
  \begin{picture}(1,0.44119268)%
    \lineheight{1}%
    \setlength\tabcolsep{0pt}%
    \put(0,0){\includegraphics[width=\unitlength,page=1]{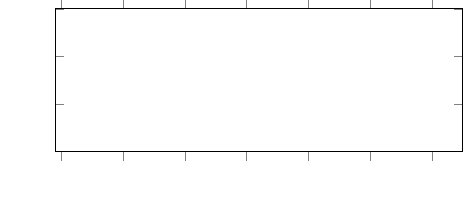}}%
    \put(0.10501221,0.05529037){\color[rgb]{0,0,0}\makebox(0,0)[lt]{\lineheight{1.25}\smash{\begin{tabular}[t]{l}255\end{tabular}}}}%
    \put(0.23823049,0.05529037){\color[rgb]{0,0,0}\makebox(0,0)[lt]{\lineheight{1.25}\smash{\begin{tabular}[t]{l}260\end{tabular}}}}%
    \put(0.37145327,0.05529037){\color[rgb]{0,0,0}\makebox(0,0)[lt]{\lineheight{1.25}\smash{\begin{tabular}[t]{l}265\end{tabular}}}}%
    \put(0.50467605,0.05529037){\color[rgb]{0,0,0}\makebox(0,0)[lt]{\lineheight{1.25}\smash{\begin{tabular}[t]{l}270\end{tabular}}}}%
    \put(0.63789883,0.05529037){\color[rgb]{0,0,0}\makebox(0,0)[lt]{\lineheight{1.25}\smash{\begin{tabular}[t]{l}275\end{tabular}}}}%
    \put(0.77111711,0.05529037){\color[rgb]{0,0,0}\makebox(0,0)[lt]{\lineheight{1.25}\smash{\begin{tabular}[t]{l}280\end{tabular}}}}%
    \put(0.90433989,0.05529037){\color[rgb]{0,0,0}\makebox(0,0)[lt]{\lineheight{1.25}\smash{\begin{tabular}[t]{l}285\end{tabular}}}}%
    \put(0.08672603,0.10141079){\color[rgb]{0,0,0}\makebox(0,0)[lt]{\lineheight{1.25}\smash{\begin{tabular}[t]{l}0\end{tabular}}}}%
    \put(0.08672603,0.20423459){\color[rgb]{0,0,0}\makebox(0,0)[lt]{\lineheight{1.25}\smash{\begin{tabular}[t]{l}5\end{tabular}}}}%
    \put(0.06796289,0.30705388){\color[rgb]{0,0,0}\makebox(0,0)[lt]{\lineheight{1.25}\smash{\begin{tabular}[t]{l}10\end{tabular}}}}%
    \put(0.06796289,0.40987318){\color[rgb]{0,0,0}\makebox(0,0)[lt]{\lineheight{1.25}\smash{\begin{tabular}[t]{l}15\end{tabular}}}}%
    \put(0,0){\includegraphics[width=\unitlength,page=2]{data/histogram.pdf}}%
    \put(0.43051891,0.0005713){\color[rgb]{0,0,0}\makebox(0,0)[lt]{\lineheight{1.25}\smash{\begin{tabular}[t]{l}IBL mean in ms\end{tabular}}}}%
    \put(0.03065259,0.1114313){\color[rgb]{0,0,0}\rotatebox{90}{\makebox(0,0)[lt]{\lineheight{1.25}\smash{\begin{tabular}[t]{l}pseudonyms in bin\end{tabular}}}}}%
  \end{picture}%
  \endgroup
  \caption{Histogram of observations in the field experiment}
  \vspace{.6cm}
  \label{fig: public}
\end{figure}

\begin{table}
  \caption{Isolated \ac{IBL} measurements of 15 smartphones}
  \label{table}
  \centering
  \scriptsize
  \sffamily
  \renewcommand{\arraystretch}{1.3}
  \setlength{\tabcolsep}{.2cm}
  \begin{tabular}{llclc}
    \hline
    \textbf{Device}					  & \textbf{OS}	 	& \textbf{Pseudonyms}			& \textbf{Mean} & \textbf{Double stdev.} \\
    \hline
    Google Pixel 4a (5G)              & Android 12 		& 10							& 286.38		& 0.41 \\
    Huawei Mate 10                    & Android 10 		& 38							& 283.04		& 0.24 \\
    Huawei P10                        & Android 9 		& 11							& 283.02    	& 0.3  \\
    Huawei P10 Lite                   & Android 8 		& 4								& 261.92		& 0.21 \\
    iPhone 13                         & iOS 15			& 3								& 274.98		& 0.19 \\
    iPhone 13 Mini\textsuperscript{a} & iOS 15			& 4								& 274.96		& 0.12 \\
    iPhone 13 Mini\textsuperscript{b} & iOS 15 			& 5								& 275.36		& 0.06 \\
    iPhone 13 Mini\textsuperscript{c} & iOS 15 			& 4								& 275.05		& 0.16 \\
    iPhone X                          & iOS 15 			& 8								& 271.74		& 0.24 \\
    OnePlus Nord 					  & Android 12 		& 28							& 286.28		& 0.2  \\
    OnePlus Nord 2 					  & Android 11 		& 7								& 270			& 0.44 \\
    Redmi Note 11 Pro 				  & Android 12 		& 9								& 286.01		& 0.67 \\
    Samsung Galaxy A51				  & Android 11 		& 7								& 286.11		& 0.31 \\
    Samsung Galaxy A6				  & Android 10 		& 3								& 283.1			& 0.1  \\
    Samsung Galaxy J7 				  & Android 9 		& 3								& 282.96		& 0.12 \\
    \hline
  \end{tabular}
\end{table}

\begin{enumerate}
  \setcounter{enumi}{2}
  \item The results from the isolated experiment are reflected in observations of public spaces.
\end{enumerate}
All means from \cref{table} also arise in the histogram of the roughly 121~observed pseudonyms in public (cf.~\cref{fig: public}).
Regarding the fact that we did not prevent phones from possibly contributing twice to our measurements---i.e. the 121 pseudonyms could potentially originate from only 110 devices---the distribution of the \acp{IBL} must be taken with caution.
However, it shows that the sample of phones in \cref{table} is not considerably different from what an adversary would observe in public spaces.

\begin{enumerate}
  \setcounter{enumi}{3}
  \item This behavior is not consistent with the \ac{GAEN} documentation which specifies an \ac{IBL} of \SIrange{200}{270}{\milli\second}~\cite{gaen-bluetooth}.
\end{enumerate}
Although this may not be crucial, it raises the questions as whether and how the specification can be adapted to improve privacy and how smartphones can be made to follow this specification.

\subsection{Quantification of Fingerprintability}

We apply fingerprinting anonymity as a privacy metric to the above data in order to quantify the information provided by the \ac{IBL}.
Therefore, we first need to determine the precision $\varepsilon$ with which an adversary can observe the \ac{IBL} mean.
As each device $d$ apparently targets the same \ac{IBL} during different pseudonym cycles, one can reasonably argue that the pseudonyms' \ac{IBL} means are normally distributed around the device's \ac{IBL} $\mu_d$.
Consequently, \SI{95}{\percent} of pseudonyms' \ac{IBL} means would lie within the range $\mu_d \pm 2\sigma_d$ of two standard deviations.
By averaging the values from \cref{table}, we obtain a \emph{precision} of
\[ \varepsilon = \frac{1}{15} \sum_{\substack{d \text{ device} \\ \text{in \cref{table}}}} 2\sigma_d = 0.251\bar 3 \approx 0.25. \]
This value determines the bin width of the histogram in the fingerprinting anonymity quantification.

The histogram shown in \cref{fig: public} is already the result of dividing the field experiment data into a histogram of bin width $\varepsilon$ with a maximal entropy of $H(X) = 4.88$.
Theoretically, these $4.88$ bits of information suffice to distinguish $2^{4.88} \approx 29$ devices.
This number must be noted cautiously for different reasons.
On the one hand, we cannot exclude the possibility that our measurements overestimate the real entropy of the \ac{IBL} which would make more devices indistinguishable than assumed.
On the other hand, a real adversary could exploit additional heuristics such as asynchronous pseudonym changes or signal strength to link pseudonyms efficiently.

The \ac{IBL} mean thus provides a fingerprinting anonymity of
\[ 1 - \frac{H(X)}{\log_2(121)} = 0.29. \]
With lower values implying less privacy protection, one might consider this result as a warning and call for a closer investigation whether users of the \ac{CWA} are exposed to a disproportional privacy risk.
However, this warning needs to come with a caveat:
Fingerprinting anonymity---like the degree of anonymity \cite{diaz2002towards} from which it is derived---should be interpreted as a relative measure which is meant to compare different scenarios.
Hence, our calculations here are merely setting a baseline for further investigations that might help fine-tuning parameters towards an optimal balance between privacy protection and utility of the \ac{CWA}.

\section{Discussion and Future Work}
\label{sec: discussion}

\ac{GAEN}-based apps such as the German \ac{CWA} turn smartphones into continuous radio wave emitters and raise questions about their users' privacy.
The privacy protection of \ac{GAEN} relies on the assumption that a smartphone's broadcasted pseudonyms cannot be linked.
If this fails to be the case, various attacks such as trajectory reconstruction could arise.
Against this background, unlinkability of randomized pseudonyms should not be taken for granted but must be ensured and verified.

Our results indicate that the temporal differences in the broadcast behavior can potentially be exploited to link pseudonyms of the \ac{CWA}.
To illustrate how an adversary could proceed, observe that the Huawei Mate~10 from \cref{table} is present in the screenshot in \cref{fig: screenshot}.
The first and last entry are clearly similar in terms of their mean and much different from all other observed pseudonyms.
Moreover, the last pseudonym in the list was observed for the first time just a few seconds after the first one stopped broadcasting.
In various scenarios, these information may be enough to link these two pseudonyms.
We quantified the information provided by the \ac{IBL} to be $4.88$ bits which is theoretically enough to distinguish $29$ devices.
As pointed out, this quantitative result is subject to some uncertainty due to the small sizes of our experiments.

It is of particular interest for future studies to investigate which factors have an influence on the \ac{IBL}.
As we did not conduct any reverse engineering, we cannot answer this question definitely but may discuss various approaches.
Overall, our observations lead us to the conjecture that a smartphone's \ac{IBL} is mostly affected by two factors:
\begin{itemize}
  \item \emph{Its hardware stack.}
        By design, the \ac{GAEN} \ac{API} frequently accesses the phone's Bluetooth hardware and is thus influenced by the physical characteristics of the device.
        For example, our experiment included three iPhone~13 Mini as well as two phones from different manufactures sharing the same chipset (the Google Pixel~4a and the OnePlus Nord have a Qualcomm Snapdragon~765G built in), and the devices exhibited similar \acp{IBL} in both cases.
  \item \emph{Its usage and multitasking.}
        Whenever two processes demand hardware resources at the same time, they are granted access by the operating system's scheduler in a specific order.
        The mentioned frequent access to processor and Bluetooth consequently causes a waiting time for the \ac{GAEN} process if the demanded resources are already allocated.
        If this waiting time has a notable influence on the \ac{IBL}, then the latter might change with a varying usage.
        During our experiments we found subtle hints that the \ac{IBL} may be slightly prolonged when another app heavily uses Bluetooth, but we could not examine this any further.
        If it turns out to be the case, an active adversary could disturb phones (e.g. the processor via network queries) and observe changes in their \acp{IBL} to gain further information about which phone broadcasts which pseudonym.
\end{itemize}
Moreover, we expect that this behavior is not limited to the German \ac{CWA} but also appears in the context of other \ac{GAEN} apps.

\section{Conclusion}

This exploratory study demonstrated that the German \ac{CWA} is vulnerable to device fingerprinting.
Smartphones with installed \ac{CWA} target a device-specific latency between two subsequent Bluetooth broadcasts.
This latency can potentially identify a smartphone, among others, and can be measured with no more than a few minutes of passive Bluetooth observation.
Contrary to public assurances, regular pseudonym changes---as implemented today---are not enough to disguise a user reliably.

Our work contributes to the costs and effectiveness of \acp{CTA} by indicating that the \ac{CWA}'s privacy impact could be higher than expected.
This becomes more significant since passive Bluetooth sniffing attacks are virtually unpreventable, and the affected \ac{OS}-level code cannot be easily removed from the users' smartphones.
Hence, any non-negligible risk of device fingerprinting needs to be considered in the evaluation and further development of \acp{CTA}.

Given that medical experts do expect the next similar pandemic soon, the time is now! We should work to reduce the fingerprintability of continuously sending \ac{BLE} devices. As a side effect, a more privacy-friendly version of pseudonym-changing protocols with \ac{BLE} or other wireless technologies might open up opportunities for other, more mundane uses of such technologies.

\bibliographystyle{ACM-Reference-Format}
\bibliography{bibliography}


\begin{thebibliography}{26}


\ifx \showCODEN    \undefined \def \showCODEN     #1{\unskip}     \fi
\ifx \showDOI      \undefined \def \showDOI       #1{#1}\fi
\ifx \showISBNx    \undefined \def \showISBNx     #1{\unskip}     \fi
\ifx \showISBNxiii \undefined \def \showISBNxiii  #1{\unskip}     \fi
\ifx \showISSN     \undefined \def \showISSN      #1{\unskip}     \fi
\ifx \showLCCN     \undefined \def \showLCCN      #1{\unskip}     \fi
\ifx \shownote     \undefined \def \shownote      #1{#1}          \fi
\ifx \showarticletitle \undefined \def \showarticletitle #1{#1}   \fi
\ifx \showURL      \undefined \def \showURL       {\relax}        \fi
\providecommand\bibfield[2]{#2}
\providecommand\bibinfo[2]{#2}
\providecommand\natexlab[1]{#1}
\providecommand\showeprint[2][]{arXiv:#2}

\bibitem[gae(2020)]%
        {gaen}
 \bibinfo{year}{2020}\natexlab{}.
\newblock \bibinfo{booktitle}{\emph{{Exposure notifications: Helping fight
  covid-19}}}.
\newblock
\urldef\tempurl%
\url{https://google.com/covid19/exposurenotifications/}
\showURL{%
\tempurl}


\bibitem[ope(2020)]%
        {open-source-project-cwa}
 \bibinfo{year}{2020}\natexlab{}.
\newblock \bibinfo{booktitle}{\emph{{Open-Source Project Corona-Warn-App}}}.
\newblock
\urldef\tempurl%
\url{https://coronawarn.app/en/}
\showURL{%
\tempurl}


\bibitem[Acar et~al\mbox{.}(2014)]%
        {acar_web_2014}
\bibfield{author}{\bibinfo{person}{Gunes Acar}, \bibinfo{person}{Christian
  Eubank}, \bibinfo{person}{Steven Englehardt}, \bibinfo{person}{Marc Juarez},
  \bibinfo{person}{Arvind Narayanan}, {and} \bibinfo{person}{Claudia Diaz}.}
  \bibinfo{year}{2014}\natexlab{}.
\newblock \showarticletitle{{The Web Never Forgets: Persistent Tracking
  Mechanisms in the Wild}}. In \bibinfo{booktitle}{\emph{{Proceedings of the
  2014 ACM SIGSAC Conference on Computer and Communications Security}}}.
  \bibinfo{pages}{674--689}.
\newblock
\urldef\tempurl%
\url{https://doi.org/10.1145/2660267.2660347}
\showDOI{\tempurl}


\bibitem[Adamsky et~al\mbox{.}(2018)]%
        {csi:adamsky:inproceeding:2018}
\bibfield{author}{\bibinfo{person}{Florian Adamsky}, \bibinfo{person}{Tatiana
  Retunskaia}, \bibinfo{person}{Stefan Schiffner}, \bibinfo{person}{Christian
  K\"{o}bel}, {and} \bibinfo{person}{Thomas Engel}.}
  \bibinfo{year}{2018}\natexlab{}.
\newblock \showarticletitle{{Poster: WLAN Device Fingerprinting Using Channel
  State Information (CSI)}}. In \bibinfo{booktitle}{\emph{{Proceedings of the
  11th ACM Conference on Security \& Privacy in Wireless and Mobile Networks}}}
  (Stockholm, Sweden) \emph{(\bibinfo{series}{WiSec '18})}.
  \bibinfo{publisher}{ACM}, \bibinfo{address}{New York, NY, USA},
  \bibinfo{pages}{277--278}.
\newblock
\showISBNx{978-1-4503-5731-9}
\urldef\tempurl%
\url{https://doi.org/10.1145/3212480.3226099}
\showDOI{\tempurl}


\bibitem[Apple and Google(2020)]%
        {gaen-bluetooth}
\bibfield{author}{\bibinfo{person}{Apple} {and} \bibinfo{person}{Google}.}
  \bibinfo{year}{2020}\natexlab{}.
\newblock \bibinfo{booktitle}{\emph{{Exposure Notification – Bluetooth
  Specification}}}.
\newblock
\urldef\tempurl%
\url{https://blog.google/documents/70/Exposure_Notification_-_Bluetooth_Specification_v1.2.2.pdf/}
\showURL{%
\tempurl}


\bibitem[{Bluetooth Special Interest Group}(2021)]%
        {bluetooth}
\bibfield{author}{\bibinfo{person}{{Bluetooth Special Interest Group}}.}
  \bibinfo{year}{2021}\natexlab{}.
\newblock \bibinfo{booktitle}{\emph{{Bluetooth Core Specification v5.3}}}.
\newblock
\urldef\tempurl%
\url{https://www.bluetooth.com/specifications/specs/core-specification-5-3/}
\showURL{%
\tempurl}


\bibitem[Cao et~al\mbox{.}(2017)]%
        {webgl:cao:2017:inproceedings}
\bibfield{author}{\bibinfo{person}{Yinzhi Cao}, \bibinfo{person}{Song Li},
  {and} \bibinfo{person}{Erik Wijmans}.} \bibinfo{year}{2017}\natexlab{}.
\newblock \showarticletitle{{(Cross-)Browser Fingerprinting via OS and Hardware
  Level Features}}. In \bibinfo{booktitle}{\emph{Proceedings of the Network and
  Distributed System Security Symposium (NDSS) 2017}}.
\newblock
\urldef\tempurl%
\url{https://doi.org/10.14722/ndss.2017.23152}
\showDOI{\tempurl}


\bibitem[Celosia and Cunche(2019)]%
        {celosia2019fingerprinting}
\bibfield{author}{\bibinfo{person}{Guillaume Celosia} {and}
  \bibinfo{person}{Mathieu Cunche}.} \bibinfo{year}{2019}\natexlab{}.
\newblock \showarticletitle{{Fingerprinting bluetooth-low-energy devices based
  on the generic attribute profile}}. In \bibinfo{booktitle}{\emph{{Proceedings
  of the 2\textsuperscript{nd} International ACM Workshop on Security and
  Privacy for the Internet-of-Things}}}. \bibinfo{pages}{24--31}.
\newblock


\bibitem[D{\'i}az et~al\mbox{.}(2003)]%
        {diaz2002towards}
\bibfield{author}{\bibinfo{person}{Claudia D{\'i}az}, \bibinfo{person}{Stefaan
  Seys}, \bibinfo{person}{Joris Claessens}, {and} \bibinfo{person}{Bart
  Preneel}.} \bibinfo{year}{2003}\natexlab{}.
\newblock \showarticletitle{{Towards measuring anonymity}}. In
  \bibinfo{booktitle}{\emph{{Privacy Enhancing Technologies}}}.
  \bibinfo{publisher}{Springer Berlin Heidelberg}, \bibinfo{pages}{54--68}.
\newblock
\urldef\tempurl%
\url{https://doi.org/10.1007/3-540-36467-6_5}
\showDOI{\tempurl}


\bibitem[Eckersley(2010)]%
        {panopticlick:eckersley:inproceedings:2010}
\bibfield{author}{\bibinfo{person}{Peter Eckersley}.}
  \bibinfo{year}{2010}\natexlab{}.
\newblock \showarticletitle{{How Unique Is Your Web Browser?}}. In
  \bibinfo{booktitle}{\emph{{Proceedings of the 10th Privacy Enhancing
  Technologies Symposium ({PETS} 2010)}}} (Berlin, Heidelberg).
  \bibinfo{publisher}{Springer Berlin Heidelberg}, \bibinfo{pages}{1--18}.
\newblock
\urldef\tempurl%
\url{https://doi.org/10.1007/978-3-642-14527-8\_1}
\showDOI{\tempurl}


\bibitem[{European Union}(2016)]%
        {gdpr}
\bibfield{author}{\bibinfo{person}{{European Union}}.}
  \bibinfo{year}{2016}\natexlab{}.
\newblock \bibinfo{title}{{Regulation (EU) 2016/679 of the European Parliament
  and of the Council of 27 April 2016 on the protection of natural persons with
  regard to the processing of personal data and on the free movement of such
  data, and repealing Directive 95/46/EC (General Data Protection
  Regulation)}}.
\newblock
\newblock
\urldef\tempurl%
\url{https://eur-lex.europa.eu/eli/reg/2016/679/oj}
\showURL{%
\tempurl}


\bibitem[Frolov and Wustrow(2019)]%
        {frolov_use_2019}
\bibfield{author}{\bibinfo{person}{Sergey Frolov} {and} \bibinfo{person}{Eric
  Wustrow}.} \bibinfo{year}{2019}\natexlab{}.
\newblock \showarticletitle{{The use of {TLS} in Censorship Circumvention}}. In
  \bibinfo{booktitle}{\emph{{Proceedings 2019 Network and Distributed System
  Security Symposium (NDSS)}}}. \bibinfo{publisher}{Internet Society}.
\newblock
\urldef\tempurl%
\url{https://doi.org/10.14722/ndss.2019.23511}
\showDOI{\tempurl}


\bibitem[He et~al\mbox{.}(2020)]%
        {he2020temporal}
\bibfield{author}{\bibinfo{person}{Xi He}, \bibinfo{person}{Eric~HY Lau},
  \bibinfo{person}{Peng Wu}, \bibinfo{person}{Xilong Deng},
  \bibinfo{person}{Jian Wang}, \bibinfo{person}{Xinxin Hao},
  \bibinfo{person}{Yiu~Chung Lau}, \bibinfo{person}{Jessica~Y Wong},
  \bibinfo{person}{Yujuan Guan}, \bibinfo{person}{Xinghua Tan},
  {et~al\mbox{.}}} \bibinfo{year}{2020}\natexlab{}.
\newblock \showarticletitle{{Temporal dynamics in viral shedding and
  transmissibility of COVID-19}}.
\newblock \bibinfo{journal}{\emph{Nature medicine}} \bibinfo{volume}{26},
  \bibinfo{number}{5} (\bibinfo{year}{2020}), \bibinfo{pages}{672--675}.
\newblock
\urldef\tempurl%
\url{https://doi.org/10.1038/s41591-020-0869-5}
\showDOI{\tempurl}


\bibitem[Hua et~al\mbox{.}(2018)]%
        {csi:hua:inproceeding:2018}
\bibfield{author}{\bibinfo{person}{Jingyu Hua}, \bibinfo{person}{Mr~Hongyi
  Sun}, \bibinfo{person}{Mr~Zhenyu Shen}, \bibinfo{person}{Zhiyun Qian}, {and}
  \bibinfo{person}{Dr~Sheng Zhong}.} \bibinfo{year}{2018}\natexlab{}.
\newblock \showarticletitle{{Accurate and Efficient Wireless Device
  Fingerprinting Using Channel State Information}}. In
  \bibinfo{booktitle}{\emph{{Proceedings of the IEEE International Conference
  on Computer Communications (INFOCOM)}}}. \bibinfo{pages}{9}.
\newblock


\bibitem[Huang et~al\mbox{.}(2014)]%
        {huang2014blueid}
\bibfield{author}{\bibinfo{person}{Jun Huang}, \bibinfo{person}{Wahhab
  Albazrqaoe}, {and} \bibinfo{person}{Guoliang Xing}.}
  \bibinfo{year}{2014}\natexlab{}.
\newblock \showarticletitle{{BlueID: A practical system for Bluetooth device
  identification}}. In \bibinfo{booktitle}{\emph{{IEEE INFOCOM 2014-IEEE
  Conference on Computer Communications}}}. IEEE, \bibinfo{pages}{2849--2857}.
\newblock


\bibitem[Husák et~al\mbox{.}(2016)]%
        {husak_https_2016}
\bibfield{author}{\bibinfo{person}{Martin Husák}, \bibinfo{person}{Milan
  Čermák}, \bibinfo{person}{Tomáš Jirsík}, {and} \bibinfo{person}{Pavel
  Čeleda}.} \bibinfo{year}{2016}\natexlab{}.
\newblock \showarticletitle{{HTTPS traffic analysis and client identification
  using passive SSL/TLS fingerprinting}}.
\newblock  \bibinfo{volume}{2016}, \bibinfo{number}{1} (\bibinfo{year}{2016}),
  \bibinfo{pages}{6}.
\newblock
\showISSN{1687-417X}
\urldef\tempurl%
\url{https://doi.org/10.1186/s13635-016-0030-7}
\showDOI{\tempurl}


\bibitem[Jana and Kasera(2009)]%
        {jana2008fast}
\bibfield{author}{\bibinfo{person}{Suman Jana} {and}
  \bibinfo{person}{Sneha~Kumar Kasera}.} \bibinfo{year}{2009}\natexlab{}.
\newblock \showarticletitle{{On Fast and Accurate Detection of Unauthorized
  Wireless Access Points Using Clock Skews}}. In
  \bibinfo{booktitle}{\emph{{Proceedings of the 14\textsuperscript{th} ACM
  international conference on Mobile computing and networking}}}.
  \bibinfo{pages}{104--115}.
\newblock
\urldef\tempurl%
\url{https://doi.org/10.1109/TMC.2009.145}
\showDOI{\tempurl}


\bibitem[Kohno et~al\mbox{.}(2005)]%
        {kohno2005remote}
\bibfield{author}{\bibinfo{person}{Tadayoshi Kohno}, \bibinfo{person}{Andre
  Broido}, {and} \bibinfo{person}{Kimberly~C Claffy}.}
  \bibinfo{year}{2005}\natexlab{}.
\newblock \showarticletitle{{Remote physical device fingerprinting}}.
\newblock \bibinfo{journal}{\emph{IEEE Transactions on Dependable and Secure
  Computing}} \bibinfo{volume}{2}, \bibinfo{number}{2} (\bibinfo{year}{2005}),
  \bibinfo{pages}{93--108}.
\newblock
\urldef\tempurl%
\url{https://doi.org/10.1109/TDSC.2005.26}
\showDOI{\tempurl}


\bibitem[Laperdrix et~al\mbox{.}(2019)]%
        {browser-fp-survey:laperdrix:article:2019}
\bibfield{author}{\bibinfo{person}{Pierre Laperdrix}, \bibinfo{person}{Nataliia
  Bielova}, \bibinfo{person}{Benoit Baudry}, {and} \bibinfo{person}{Gildas
  Avoine}.} \bibinfo{year}{2019}\natexlab{}.
\newblock \showarticletitle{{Browser Fingerprinting: A survey}}.
\newblock  (\bibinfo{year}{2019}).
\newblock
\showeprint[arxiv]{1905.01051}
\urldef\tempurl%
\url{http://arxiv.org/abs/1905.01051}
\showURL{%
\tempurl}


\bibitem[Mayer(2009)]%
        {anonymity:mayer:thesis:2009}
\bibfield{author}{\bibinfo{person}{Jonathan~R Mayer}.}
  \bibinfo{year}{2009}\natexlab{}.
\newblock \bibinfo{title}{{``Any person... a pamphleteer:'' Internet Anonymity
  in the Age of Web 2.0}}.
\newblock
\newblock
\newblock
\shownote{Bachelor Thesis}.


\bibitem[Mowery and Shacham(2012)]%
        {mowery_pixel_2012}
\bibfield{author}{\bibinfo{person}{Keaton Mowery} {and} \bibinfo{person}{Hovav
  Shacham}.} \bibinfo{year}{2012}\natexlab{}.
\newblock \showarticletitle{{Pixel Perfect: Fingerprinting Canvas in HTML5}}.
  In \bibinfo{booktitle}{\emph{{Proceedings of W2SP 2012}}}.
  \bibinfo{pages}{12}.
\newblock


\bibitem[Pfitzmann and Hansen(2010)]%
        {pfitzmann2010terminology}
\bibfield{author}{\bibinfo{person}{Andreas Pfitzmann} {and}
  \bibinfo{person}{Marit Hansen}.} \bibinfo{year}{2010}\natexlab{}.
\newblock \bibinfo{title}{{A terminology for talking about privacy by data
  minimization: Anonymity, unlinkability, undetectability, unobservability,
  pseudonymity, and identity management}}.
\newblock
\newblock


\bibitem[Prodan et~al\mbox{.}(2022)]%
        {Prodan_Birov_2022}
\bibfield{author}{\bibinfo{person}{Alexandra Prodan}, \bibinfo{person}{Strahil
  Birov}, \bibinfo{person}{Viktor von Wyl}, {and} \bibinfo{person}{Wolfgang
  Ebbers}.} \bibinfo{year}{2022}\natexlab{}.
\newblock \bibinfo{booktitle}{\emph{{Digital Contact Tracing Study --- Study on
  lessons learned, best practices and epidemiological impact of the common
  European approach on digital contact tracing to combat and exit the COVID-19
  pandemic}}}.
\newblock \bibinfo{publisher}{European Commission}.
\newblock
\showISBNx{978-92-76-58985-3}


\bibitem[Sit(2017)]%
        {leen_mimo_2017}
\bibfield{author}{\bibinfo{person}{Yoke~Leen Sit}.}
  \bibinfo{year}{2017}\natexlab{}.
\newblock \bibinfo{booktitle}{\emph{{MIMO} {OFDM} Radar-Communication System
  with Mutual Interference Cancellation}}.
\newblock \bibinfo{publisher}{{KIT} Scientific Publishing}.
\newblock
\showISBNx{978-3-7315-0599-0}


\bibitem[Van~Kerkhove et~al\mbox{.}(2021)]%
        {van2021preparing}
\bibfield{author}{\bibinfo{person}{Maria~D Van~Kerkhove},
  \bibinfo{person}{Michael~J Ryan}, {and} \bibinfo{person}{Tedros~Adhanom
  Ghebreyesus}.} \bibinfo{year}{2021}\natexlab{}.
\newblock \showarticletitle{Preparing for “Disease X”}.
\newblock \bibinfo{journal}{\emph{Science}} \bibinfo{volume}{374},
  \bibinfo{number}{6566} (\bibinfo{year}{2021}), \bibinfo{pages}{377}.
\newblock


\bibitem[Xue et~al\mbox{.}(2022)]%
        {280012}
\bibfield{author}{\bibinfo{person}{Diwen Xue}, \bibinfo{person}{Reethika
  Ramesh}, \bibinfo{person}{Arham Jain}, \bibinfo{person}{Michalis Kallitsis},
  \bibinfo{person}{J.~Alex Halderman}, \bibinfo{person}{Jedidiah~R. Crandall},
  {and} \bibinfo{person}{Roya Ensafi}.} \bibinfo{year}{2022}\natexlab{}.
\newblock \showarticletitle{{OpenVPN is Open to VPN Fingerprinting}}. In
  \bibinfo{booktitle}{\emph{{31\textsuperscript{st} USENIX Security Symposium
  (USENIX Security 22)}}}. \bibinfo{publisher}{USENIX Association},
  \bibinfo{address}{Boston, MA}, \bibinfo{pages}{483--500}.
\newblock
\showISBNx{978-1-939133-31-1}


\end{thebibliography}

\end{document}